\documentclass[prx,superscriptaddress,amsfonts,amssymb,amsmath,twocolumn]{revtex4}
    \usepackage{epsfig}
    \usepackage{graphics}  
    \usepackage{color}
    \date{\today}

\begin{document}

%%%%%%%%%%%%%%%%%%%%%%%%%%%%%%%%%%%%%%%%%%%%%%%%%%%%%%%%%%%%%%%%%%%%%
    %%%%%%%%%%%%%%%%%%%%%         Title      %%%%%%%%%%%%%%%%%%%%%%%%%%%
%%%%%%%%%%%%%%%%%%%%%%%%%%%%%%%%%%%%%%%%%%%%%%%%%%%%%%%%%%%%%%%%%%%%%

    \title{Proposal for observing the Unruh effect with classical electrodynamics}

%%%%%%%%%%%%%%%%%%%%%%%%%%%%%%%%%%%%%%%%%%%%%%%%%%%%%%%%%%%%%%%%%%%%%
    %%%%%%%%%%%%%%%%%%%%     Authors & Addresses %%%%%%%%%%%%%%%%%%%%%%%
%%%%%%%%%%%%%%%%%%%%%%%%%%%%%%%%%%%%%%%%%%%%%%%%%%%%%%%%%%%%%%%%%%%%%

    \author{Gabriel Cozzella}\email{cozzella@ift.unesp.br}
    \affiliation{Instituto de F\'\i sica Te\'orica, Universidade
    Estadual Paulista, Rua Dr.\ Bento Teobaldo Ferraz, 271, 01140-070,
    S\~ao Paulo, S\~ao Paulo, Brazil}

    \author{Andr\'e G.\ S.\ Landulfo}\email{andre.landulfo@ufabc.edu.br}
    \affiliation{Centro de Ci\^encias Naturais e Humanas,
    Universidade Federal do ABC,
    Avenida dos Estados, 5001, 09210-580,
    Santo Andr\'e, S\~ao Paulo, Brazil}

    \author{George E.\ A.\ Matsas}\email{matsas@ift.unesp.br}
    \affiliation{Instituto de F\'\i sica Te\'orica, Universidade
    Estadual Paulista, Rua Dr.\ Bento Teobaldo Ferraz, 271, 01140-070,
    S\~ao Paulo, S\~ao Paulo, Brazil}

    \author{Daniel A.\ T.\ Vanzella}\email{vanzella@ifsc.usp.br}
    \affiliation{Instituto de F\'\i sica de S\~ao Carlos,
    Universidade de S\~ao Paulo, Caixa Postal 369, 13560-970,
    S\~ao Carlos, S\~ao Paulo, Brazil}

%%%%%%%%%%%%%%%%%%%%%%%%%%%%%%%%%%%%%%%%%%%%%%%%%%
%%%%%%%%%%%%%%%%%%%%%           Abstract           %%%%%%%%%%%%%%%%%%%
%%%%%%%%%%%%%%%%%%%%%%%%%%%%%%%%%%%%%%%%%%%%%%%%%%

  \begin{abstract}
Although the Unruh effect can be rigorously considered as well tested as free quantum field theory itself, 
it would be nice to provide an experimental evidence of its existence. This is not easy because the linear 
acceleration needed to reach a temperature $1~{\rm K}$ is of order $10^{20}~{\rm m/s}^2$. Here, we 
propose a simple experiment reachable under present technology whose result may be directly interpreted in 
terms of the Unruh thermal bath. Instead of waiting for experimentalists to perform it, we use standard 
classical electrodynamics to anticipate its output and fulfill our goal.  
 \end{abstract}

    \maketitle

    {\bf Introduction:} In 1976 Unruh unveiled one of the most interesting 
    effects of quantum field theory according to which linearly accelerated
    observers with proper acceleration $a={\rm constant}$ in the Minkowski
    vacuum (i.e., no-particle state for inertial observers) detect a
    thermal bath of 
    particles at a temperature~\cite{U76}
    (see also note~\cite{note1}) 
    \begin{equation}
    T_{\rm U}= a \hbar / (2 \pi k_{\rm B} c).
    \label{UnruhTemperature}
    \end{equation}
    This was the completion of Fulling's discovery that inertial
    and uniformly accelerated ({\em Rindler}) observers would extract distinct
    particle contents from the same field theory~\cite{F73} and came to 
    clarify Davies' 1975 result~\cite{D75}. The rather nonintuitive
    content carried by the Unruh effect,
  %   (for a curious historical note, see \cite{note2}), 
     namely, that inertial observers in Minkowski vacuum 
    would freeze at $0~{\rm K}$ while accelerated ones would burn at high enough
    proper accelerations, was missed at first by many, including
    Bisognano and Wichmann, who obtained it
    independently~\cite{BW76} but seemingly did not realize it up to 1982,
    when Sewell connected their theorem to the Unruh
    effect~\cite{S82}. By 1984 (after the publication of Unruh and
    Wald's Ref.~\cite{UW84}), it should have become clear that the Unruh effect
    is necessary to keep the consistency of field theory in uniformly accelerated 
    frames and does not require any more experimental confirmation than
    free quantum field theory does. But sporadic claims that the Unruh
    effect  does not exist or, more often, lacks observational confirmation
    have motivated the quest for experimental evidences which could settle the 
    issue. This is not easy, however, because the linear acceleration needed to 
    reach a temperature $1~{\rm K}$ is of order $10^{20}~{\rm m/s}^2$~\cite{CHM08,FM14}. 
    Bell and Leinaas were the first to go into this by trying to explain
    the electron depolarization in storage rings in terms of the
    Unruh effect~\cite{BL83}. They achieved partial success  because
    the Unruh effect is derived for uniformly accelerated observers
    who are associated with a time-translation symmetry,
    namely, the
    boost isometry, rather than for circularly moving
    observers who cannot be connected to any analogous global 
    time-translation symmetry.
    Another proposal relied on the decay of accelerated
    protons~\cite{M97}. It was shown that Rindler observers need
    the Unruh effect to understand the decay of uniformly accelerated 
    protons~\cite{VM01}-\cite{S03}. Unfortunately (for us -- particle physicists 
    may disagree), the proton lifetime in actual accelerators is too long, 
    rendering this observation (on Earth) virtually impossible~\cite{VM01b}.
    Under typical accelerations at the LHC/CERN, the proton lifetime
    would be $10^{3 \times 10^8}~{\rm yr}$! A more promising strategy
    consists of seeking for fingerprints of the Unruh effect in the
    radiation emitted by accelerated charges. Accelerated charges
    should back react due to radiation emission, quivering
    accordingly. Such a quivering would be naturally interpreted by
    Rindler observers as a consequence of the charge interaction with
    the photons of the Unruh thermal bath~\cite{CT99}-\cite{OYZ16}. The
    scattering of Rindler photons by the charge in the accelerated
    frame would correspond in the inertial frame to the
    emission of pairs of correlated photons~\cite{SSH06}. The
    observation of such a signal could be assigned to
    the existence of the Unruh thermal bath. The difficulty with
    these proposals lies on the dependence on ultraintense lasers
    and they have never been realized.
    It happens, however, that the usual Larmor radiation which does
    not require paramount accelerations, for it is related to the
    emission probability of single photons, is already enough to
    unveil the existence of the Unruh effect as follows~\cite{HMS92}:
    each photon emitted by a uniformly accelerated charge, as
    described by inertial observers, corresponds to either the
    emission or absorption of a {\em zero-energy} Rindler photon
    to or from the Unruh thermal bath, respectively. Thus,
    the very observation of the Larmor radiation can be
    seen as a signal of the Unruh effect. The fact that a quantum
    effect [note the $\hbar$ in Eq.~(\ref{UnruhTemperature})] may
    be verified through a classical phenomenon might sound strange
    at first but there is no reason for preoccupation once one notes
    that the $\hbar$ in the thermal factor      
    $$
    \exp \left[\hbar \omega_R/(k_{\rm B} T_{\rm U}) \right] =
    \exp \left(2 \pi \omega_R c/a \right),
    $$
    associated with the Unruh thermal bath of Rindler particles
    with energy $\hbar \omega_R$, cancels out (see Ref.~\cite{HM93} 
    for a comprehensive discussion).
    For some reason, however -- perhaps because the reasoning above involves
    the unfamiliar concept of zero-energy particles
    or because the calculation required a certain regularization --,
    this result did not turn out convincing enough to settle the issue
    and papers disputing the existence of the Unruh effect can
    still be seen (see, e.g., Ref.~\cite{CM16} and references therein). 
    
    In the present paper we suggest a simple
    laboratory experiment which should be enough to make it clear that
    {\em the Unruh effect lives among us}. The idea is to consider a phenomenon
    as simple and technically feasible as in Ref.~\cite{HMS92} and, at
    the same time, free of unfamiliar concepts and technical subtleties,
    thus avoiding unnecessary concerns.
    In order to make our strategy clear, we state the experiment
    in the uniformly accelerated frame and analyze it assuming we
    are Rindler observers immersed in 
    a thermal bath of Rindler particles with temperature $T$. Then,
    we use our results to guide experimentalists (in inertial laboratories)
    about what they should seek to allege the observation of the Unruh effect.
    According to the Unruh effect, $T$ must equal 
    $T_{\rm U}$ given in Eq.~(\ref{UnruhTemperature}) but we shall
    leave $T$ as a free parameter to be {\it measured} by the inertial experimentalists
    by fitting the data. However, rather than sitting back and waiting for experimentalists to confirm 
    the prediction $T=T_{\rm U}$,
    we proceed to a straightforward calculation in the inertial
    frame, using {\it standard electrodynamics},  to confirm it by ourselves.  This must be seen 
    as a virtual observation of the Unruh effect unless one doubts standard
    electrodynamics.
    
   We adopt metric signature $(+,-,-,-)$ and units where 
    $G=c=k_{\rm B}=1$, unless stated otherwise.
  
    {\bf The physical problem:} The goal  posed by Rindler observers
    will be to calculate the photon emission rate  from a circularly moving
    charge with constant angular velocity as defined by them, assuming that 
    the electromagnetic (radiation) field is in the Minkowski vacuum, $|0_{\rm M} \rangle$, 
    which they perceive as a thermal state due to the Unruh effect. Our 
    Rindler observers will be chosen to be a congruence at 
    the (right) Rindler wedge, i.e., the $ z >|t|$ portion of the
    Minkowski spacetime, where $(t,z,r,\phi)$ are the usual
    cylindrical coordinates. By covering the Rindler wedge with 
    $(\lambda, \xi, r, \phi)$ coordinates, the line element can be written as
    $
    ds^2 = e^{2 a \xi} \left( d\lambda^2 - d\xi^2 \right) - dr^2 - r^2 d \phi^2, \label{ds2}
    $
    where $\lambda, \xi \in (-\infty, +\infty)$ are given by
    \begin{equation}
        t=a^{-1}e^{a \xi} \sinh{a\lambda},\;\;\; 
        z=a^{-1}e^{a \xi} \cosh{a\lambda}, 
    \label{Rindler}
    \end{equation}
    and $a={\rm constant}>0$.
    Each Rindler observer will be labeled by constant values of $\xi$, $r$, and $\phi$ with corresponding
    proper acceleration $a e^{-a\xi}$.

    A circularly moving charge $q$ with mass $m_q$ and worldline  $\xi =0$,
    $r=R$, and $\phi= \Omega \lambda$, with $R,\Omega = { \rm constants}$, has 4-velocity components
    \begin{equation}
    u^\mu =  \gamma \left(1,0,0,\Omega \right),
    \;\;\; \gamma = 1/\sqrt{1-R^2 \Omega^2},
    \label{4-velocityR}
    \end{equation}
    giving rise to the electric 4-current
    \begin{equation}
    j^\mu =  \frac{q \,u^\mu}{R \, u^0} \delta(\xi) \delta(\phi - \Omega \lambda) \delta(r-R).
    \label{current}
    \end{equation}
    Thus, the only free parameters are
    \begin{equation}
    R, \Omega,\, {\rm and}\; a,
    \label{free}
    \end{equation}
    where $a$ is the proper acceleration of the Rindler observers at the plane $\xi=0$.

    The Lagrangian density of the electromagnetic field $A_\mu$ willl be 
    $ {\cal L}= -\sqrt{-g}\left[ (1/4)F_{\mu \nu} F^{\mu \nu} + (2\alpha)^{-1} (\nabla_\mu A^\mu)^2 \right]$,
    leading to the following field equations~\cite{HMS92}:
    \begin{equation}
    \nabla_\mu \nabla^\mu A_\nu=0
    \label{Maxwell equations}
    \end{equation}
    in the Feynman gauge, $\alpha=1$. The four independent solutions
    $A_\mu^{(\epsilon,m,k_\bot,\omega_R)}$ of
    Eq.~(\ref{Maxwell equations}) comprise the two physical modes labeled
    by $\epsilon=1,2$ and the pure gauge and nonphysical ones labeled
    by $\epsilon=3$  and 4, respectively, with $k_\bot, \omega_R \in [0,+\infty)$,
    and $m \in \mathbb{Z}$ being the remaining quantum numbers.

    The physical modes, which must satisfy
    the Lorenz condition $\nabla^\mu A_\mu =0$ and not be pure gauge, are
    \begin{align}
    A_\mu^{(1,m,k_\bot,\omega_R)} 
    &= 
    k_\bot^{-1} 
    \left(
    \partial_\xi f_{m k_\bot \omega_R}, 
    \partial_\lambda f_{m k_\bot \omega_R} ,
    0,
    0 \right), 
    \label{physical mode 1}\\
    A_\mu^{(2,m,k_\bot,\omega_R)} 
    &= 
    k_\bot^{-1} 
    \left(
    0,
    0, 
    -m f_{m k_\bot \omega_R  }/r, 
    -i r\partial_r f_{m k_\bot \omega_R  } 
    \right), 
    \label{physical mode 2}
    \end{align}
    where
   $$
    f_{m k_\bot \omega_R} = 
    C_{\omega_R} K_{i \omega_R/a} \left( k_\bot e^{a \xi}/a \right) 
    J_m (k_\bot r ) e^{i m \phi} e^{-i \omega_R \lambda}
   $$
    satisfies
    $
      \left[ 
      e^{-2a\xi } ( \partial^2_\lambda  - \partial^2_\xi ) - \nabla_\bot^2
      \right] 
      f_{m k_\bot \omega_R} = 0
    $
    with $\nabla_\bot^2 \equiv r^{-1} \partial_r (r {\partial_r} )+ r^{-2} \partial^2_\phi$ and
    we have chosen the constant $C_{\omega_R} = [ {\sinh{(\pi \omega_R / a)}}/{(2 \pi^3 a)} ]^{1/2}$ to guarantee
    that Eqs.~(\ref{physical mode 1}) and~(\ref{physical mode 2}) are properly Klein-Gordon
    orthonormalized. (We recall that pure gauge and nonphysical modes can be chosen
    orthogonal to the physical ones.)

    Let us define the electromagnetic field operator as
    \begin{equation}
    \hat A_\mu
    =
    \sqrt \hbar
    \!\!
    \sum_{m={-\infty}}^{\infty} 
    \int_0^{\infty} \!\! d k_\bot k_\bot
    \!\!
    \int_0^{\infty} \!\! d \omega_R
     \sum_{\epsilon = 1}^{4}
    \left\{ \hat a^{\rm R}_{(i)}
           A_\mu^{(i)}
            + {\rm H.c.} \right\},
    \label{expansion1}
    \end{equation}
    where we have used the shortcut $(i) \equiv (\epsilon , m, k_\bot, \omega_R)$. The
    annihilation $\hat{a}^{\rm R}_{(i)}$ and creation $\hat{a}^{{\rm R} \dagger}_{(i)}$
    operators satisfy
        $
        [\hat{a}^{\rm R}_{(i)},\hat{a}^{{\rm R} \dagger}_{(i')}] = 
        (\hbar/ k_\bot) \delta_{\epsilon \epsilon'} 
        \delta_{m m'}\delta(k_\bot-k'_\bot) \delta(\omega_R-\omega_R')
        $
    and
    $
    [\hat{a}^{\rm R}_{(i)},\hat{a}^{\rm R}_{(i')}] = 
    [\hat{a}^{{\rm R} \dagger}_{(i)}, \hat{a}^{{\rm R} \dagger}_{(i')}] = 0$
    for physical modes $\epsilon=1,2
   $.
    The electromagnetic field is coupled to the current through the interaction
    Lagrangian density ${\cal{L}}_{\rm int} = \sqrt{-g} \ j^\mu \hat{A}_\mu$.
    The current will couple to both physical polarizations.
    The  emission and absorption photon number distribution for fixed $m$
    and transverse ``momentum'' (wave number) $k_\bot$, per Rindler observers' proper time interval 
    $\Delta \tau_{\rm  R}$, at the tree level, are
    \begin{eqnarray}
    &&\frac{1}{k_{\bot}}\frac{d\Gamma_{k_\bot m}^{{\rm R}\, {\rm em}}}{dk_\bot} 
    \equiv
    \frac{k_{\bot}^{-1}}{\Delta \tau_{\rm R}} \frac{dN_{k_\bot m}^{{\rm R}\, {\rm em}}}{dk_\bot} 
    \nonumber \\
    &=&
    \sum_{\epsilon = 1,2}
    \int_0^{\infty} d \omega_R 
    \frac{| {\cal A}^{{\rm R}\, {\rm em}}_{(i)}|^2}{\Delta \tau_{\rm R}}
    \left( 1 +\frac{1}{e^{\hbar\omega_R/T}-1} \right)
    \label{emissionRrate}
    \end{eqnarray}
    and
    \begin{eqnarray}
     &&\frac{1}{k_{\bot}} \frac{d\Gamma_{k_\bot m}^{{\rm R}\, {\rm abs}}}{dk_\bot}  
     \equiv
     \frac{k_{\bot}^{-1}}{\Delta \tau_{ R}} \frac{d N_{k_\bot m}^{{\rm R}\, {\rm abs}}}{dk_\bot} 
     \nonumber \\
    &=&
    \sum_{\epsilon = 1,2}
    \int_0^{\infty} d \omega_R 
    \frac{| {\cal A}^{{\rm R}\, {\rm abs}}_{(i)}|^2}{\Delta \tau_{\rm R}}
    \frac{1}{e^{\hbar \omega_R/T}-1}
    \label{absorptionRrate}
    \end{eqnarray}
    with
    $
       | {\cal A}^{{\rm R}\, {\rm em}}_{(i)} | =  | {\cal A}^{{\rm R}\, {\rm abs}}_{(i)} |
        = 
       \hbar^{-1}
        \left| 
        \int d^4 x \sqrt{-g}  j^\mu\, \!\langle 0_{\rm R}  | \hat{A}_\mu | i \rangle
        \right| \propto \delta(\omega_R-m\Omega)
    $
    and the Bose-Einstein thermal factors  in  Eqs.~(\ref{emissionRrate}) 
    and~(\ref{absorptionRrate})  are present  
    because of the thermal bath in the accelerated frame. However, instead of setting 
   $T=T_{\rm U}\equiv \hbar a/(2\pi)$,
    as would be enforced by the Unruh effect,
    here we leave $T$ as a free, independent parameter to be set by fitting the data measured in
    the inertial laboratory.
    (Note, from the amplitudes above, 
    that were the charge linearly accelerated, $\Omega=0$, the current would
    only interact with zero-energy Rindler photons $\omega_R=0$~\cite{HMS92}.) 
    The corresponding total 
    distribution rate (i.e., emission plus absorption) is computed to be
    \begin{eqnarray}
     \frac{d\Gamma_{k_\bot m}^{{\rm R}\, {\rm tot}}}{dk_\bot}
    &=&\frac {q^2 k_\bot}{ \pi^2  \hbar a}\Theta(m)
    \left[ |K'_{{im\Omega}/{a}} ({k_\bot}/{a})|^2 | J_m (k_\bot R ) |^2 \right.
    \nonumber \\
     &&    \left. \!\!\!\!\!\!\!\!\!\!\!
     + (R \Omega)^2 | K^{}_{{i m \Omega}/{a}} ({k_\bot}/{a} ) |^2
    | J'_m (k_\bot R) |^2  \right] \! 
   \nonumber \\
   & &\times
   \sinh \left( \frac{\pi m \Omega}{a}\right)
   \coth \left( \frac{ m \Omega \hbar}{2T}\right),
    \label{gammaRtotal}
    \end{eqnarray}
    where we have used Eqs.~(\ref{current}), (\ref{physical mode 1})-(\ref{physical mode 2}),
    and~(\ref{expansion1}) in Eqs.~(\ref{emissionRrate}) and~(\ref{absorptionRrate}), 
    $``\, ' \,  "$ means derivative with respect to the argument and 
    $\Theta (m) \equiv 0, 1/2$, and $1$ for $m<0, m=0$, and $m>0$, respectively.

    Now, Rindler observers are ready to propose a laboratory 
    experiment to be run by inertial experimentalists and predict 
    its output 
    %thanks to the Unruh effect.  
    as a function of the free parameter $T$. The confirmation of the equality $T=T_{\rm U}$
    should be seen as an as-direct-as-possible 
    verification of the Unruh effect by an inertial-lab-based experiment.

\begin{figure}
    \includegraphics[scale=0.4]{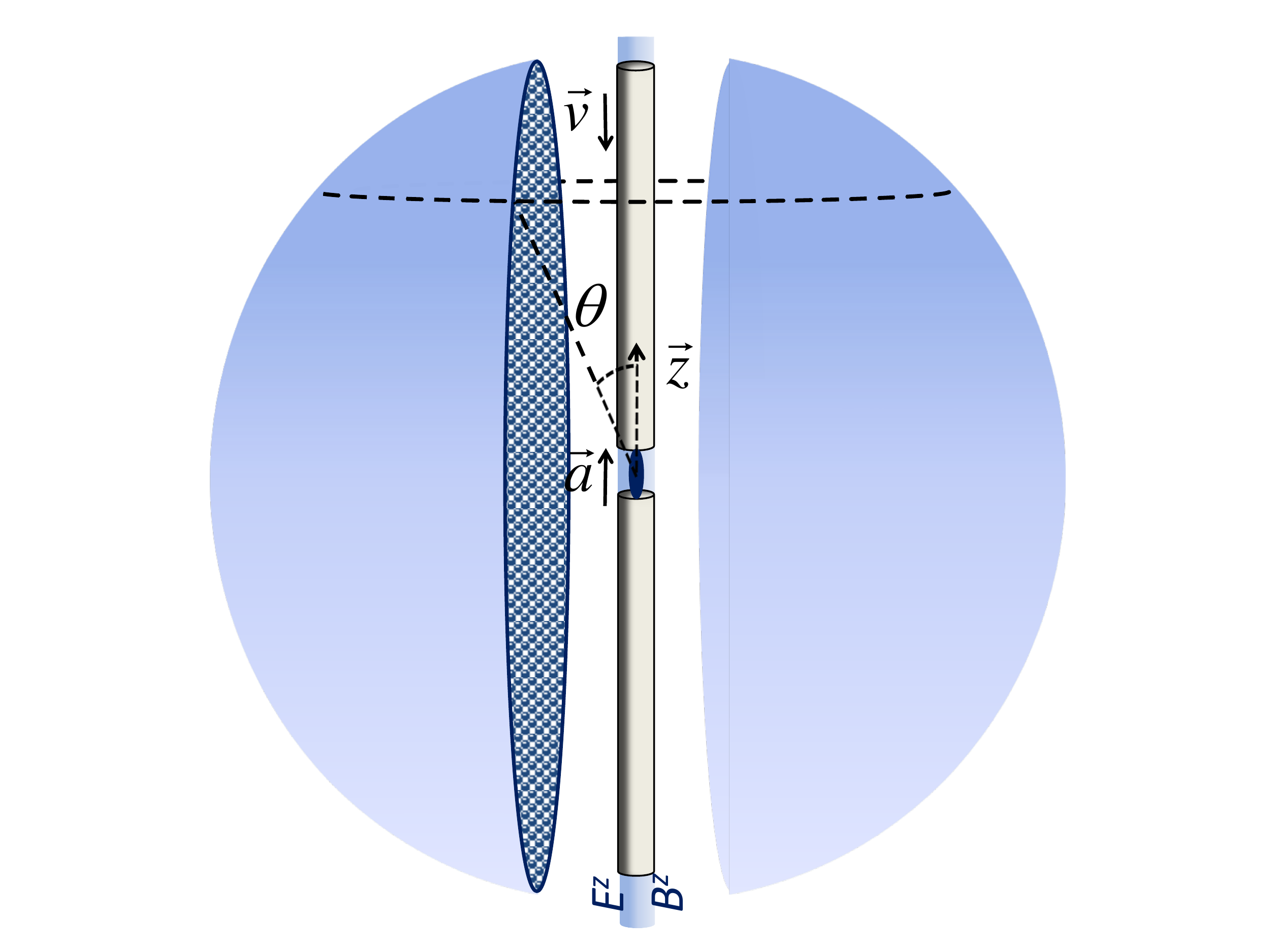}
    \caption{Electrons are injected with velocity $\vec v$ in a cylinder 
    containing suitable electric, $E^z$, and magnetic, $B^z$, fields. 
    Radiation is released near the center from an open window, which is surrounded 
    by electromagnetic detectors lying on a sphere. This allows us to obtain 
    the radiation spectral decomposition from which $dN_{k_\bot}^{{\rm M} } /d k_\bot$ 
    is calculated.}
    \label{Figure2}
    \end{figure}
    {\bf The inertial-laboratory experiment:} Let us set a pair of homogeneous and 
    constant electric, $E^z = m_q \gamma a /q$, and magnetic, $B^z =  -m_q \Omega \gamma / q$, 
    fields defined by the free parameters~(\ref{free}) along the $z$ direction. 
    Then,   a charge $q$ is injected with transverse and longitudinal velocity components, $v_{\bot}$ and $v_{\parallel}$,
    respectively, in such a way that 
    its 4-velocity $u^{\alpha}$ -- satisfying the Lorentz law of force $ m_q u^\beta \nabla_{\beta} u^\alpha= q F^{\alpha}_{\  \beta} u^\beta $ --
    is given by Eq.~(\ref{4-velocityR})~\cite{Note_velocities}. 
     In the usual cylindrical coordinates, $(t,z,r,\phi)$ with the $z$ axis 
    aligned with the 3-acceleration of the Rindler observers, the 
    Minkowski line element is
    $
    d s^2 = d t^2 - d z^2 - d r^2 - r^2 d \phi^2
    $,
    $
    u^\alpha = \gamma
            \left(
            \cosh (a \lambda),
             \sinh( a \lambda),
            0,
           \Omega
            \right)
    $, 
    and
     \begin{equation}
            F_{\alpha \beta} = \begin{pmatrix}
        0 &E^z  &0  &0 \\
         -E^z&  0&  0&0 \\
         0&  0&  0& -r B^z\\
         0&0  & r B^z &0
        \end{pmatrix}. \label{Fuv}
        \end{equation}  
    
    A prototype experimental apparatus is shown in Fig.~\ref{Figure2}, where a 
    sub-picosecond charged bunch containing $\sim 10^7$  electrons~\cite{W99}  
    is injected in a cylinder containing $E^z$ and $B^z$. The radiation released near 
    the center (where the charges are assumed to make the U turn) through an open window 
    of length $L$ is collected by detectors set on a sphere with radius $R_S \gg L$.   
    Since the charges emit radiation at typical wavelengths $\lambda \sim 1/a_{\rm tot}$, 
    where $a_{\rm tot} = \gamma^2 \sqrt{a^2+ R^2 \Omega^4}$ is the charge
   {\it total} proper acceleration,
    we should require $a_{\rm tot} \sim 1/\lambda \gg 1/L \approx 10^{17}~{\rm m/s^2} \times (1~{\rm m}/L) $ 
    in order to avoid finite-size effects coming from the window.
    We note that magnetic and electric fields $B^z \approx  10^{-1}~{\rm T}$ 
    and $E^z \approx 1~{\rm MV/m}$, respectively, achievable under present 
    technology~\cite{W08}, produce accelerations $a \sim 10^{17}~{\rm m/s^2}$ and $a_{\rm tot}\sim 10^{19}~{\rm m/s^2}$, where we 
    have assumed $R\sim 10^{-1}~{\rm m}$. We also note that the radiation backreaction 
    on the charge trajectory is negligible~\cite{note7}.    
    
    The relevant quantity to be measured by the inertial experimentalists is the spectral-angular distribution
    $$I(\omega, \theta,\phi)\equiv \frac{d{\cal E}(\omega, \theta,\phi)}{d\omega \,d(\cos\theta) \,d\phi}$$ of the emitted 
    energy ${\cal E}$. From this and the one-photon relation ${\cal E}=\hbar \omega$, we get the corresponding photon
    number:
    \begin{equation}
    dN_{\omega \theta \phi}^{{\rm M} } ={d\cal E}/ {(\hbar\omega)}= { I(\omega,\theta,\phi)}{(\hbar \omega)}^{-1} d\omega\, d(\cos\theta) \,d\phi,
    \label{N}
    \end{equation}
    which leads to the  $k_\bot$-distribution of radiated photons
    \begin{equation}
     \frac{dN_{k_\bot}^{{\rm M} } }{dk_\bot} 
     = \frac{k_\bot}{\hbar} \int_0^{2\pi} d\phi \int_{-\infty}^{\infty}  
     \frac{dk_z}{(k_\bot^2 + k_z^2)^{3/2}} I(\omega, \theta,\phi)
    \label{distribuicao inercial}
    \end{equation}
    (recall that $\omega^2 = k_{\bot}^2+k_z^2$, $k_{\bot}=\omega \sin\theta$, and
  $d\omega\, d(\cos\theta)  = \omega^{-2} k_\bot dk_z dk_\bot$).
    This is the quantity for which the uniformly accelerated observers
    can make a prediction, for, according to the Unruh effect~\cite{UW84},
{\em 
each emission of a Minkowski photon according to inertial observers
corresponds to the absorption or emission of a Rindler photon 
from or to the Unruh thermal bath, respectively}~\cite{note3}. Therefore, 
the validity of the Unruh effect demands
    \begin{eqnarray}
    & &  
    \!\! \!\!\!\!\!\!\!\!\!\!\!\! \frac{dN_{k_\bot}^{{\rm M} }}{dk_\bot} 
    \propto 
    \left.\sum_{m=-\infty}^{\infty} \frac{d\Gamma_{k_\bot m}^{{\rm R}\, {\rm tot}}}{dk_\bot}
    \right|_{T=T_{\rm U}}\nonumber \\
    & & 
    \!\! \!\!\!\!\!\!\!\!\!\!\!\!=\frac {q^2 k_\bot}{ \pi^2 \hbar a}\sum_{m=-\infty}^{\infty}\Theta(m)
    \left[ |K'_{{im\Omega}/{a}} ({k_\bot}/{a})|^2 | J_m (k_\bot R ) |^2 \right.
    \nonumber \\
     &&    \left. 
     \!\!\!\!\!\!\!\!\!\!\!\!\!\! + (R \Omega)^2 | K_{{i m \Omega}/{a}} ({k_\bot}/{a} ) |^2
    | J'_m (k_\bot R) |^2  \right]
   \cosh \left[ \frac{\pi m \Omega}{a}\right] \!\!.
    \label{Crucial}
    \end{eqnarray}
\begin{figure}
    \centering
    \includegraphics[scale=0.37]{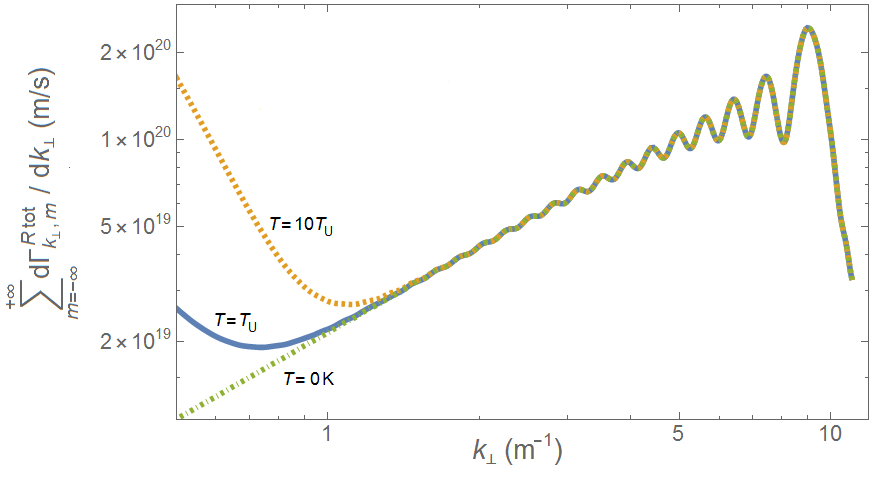}
    \caption{For the sake of illustration, we plot Eq.~(\ref{gammaRtotal}) summed in $m$ for different
    values of $T$ assuming $E^z = 1$~MV/m, $B^z = 10^{-1}$~T, $R = 10^{-1}$~m, and injection energy
    $3.5$~MeV.
    The right-hand side of Eq.~(\ref{Crucial}) corresponds to the solid-line curve.}
    \label{FigureGraph}
    \end{figure}
    The proportionality sign appears because the total number of photons 
    depends on how long the experiment is run. In Fig.~\ref{FigureGraph}, 
    we plot the right-hand side of Eq.~(\ref{gammaRtotal}) summed in $m$ 
    for different values of $T$. The prediction given in Eq.~(\ref{Crucial}) 
    is represented by the solid-line curve ($T=T_{\rm U}$). We must keep in mind 
    that due to finite-size effects coming from the window, greater experimental  
    care must be taken in the region $k_{\bot}\lesssim 1/L$.

   {\bf Virtual confirmation of the Unruh effect:} 
   Rather than waiting for experimentalists to confirm the prediction~(\ref{Crucial}), 
   here we perform a classical-electrodynamics calculation 
   to prove it -- to the extent one trusts classical electromagnetism. 
   The spectral-angular  distribution $I(\omega, \theta, \phi)$ is expressed in terms of the angular 
    distribution of the radiated electric field $\vec{E}_{\rm rad}(t,\theta,\phi)$~\cite{Z12}:
    \begin{equation}
    I(\omega, \theta,\phi) 
    = \frac{R_S^2}{\pi} \left|\int_{-\infty}^{\infty} 
    d t \ \vec{E}_{rad}(t,\theta,\phi) e^{-i \omega t} \right|^2. \label{SpecDenERad}
    \end{equation}
   For 
   our accelerated point-like charge, 
   Eq.~(\ref{SpecDenERad}) can be written as (see, e.g., Ref.~\cite{Z12})
   \begin{equation}
     I(\omega, \theta,\phi) 
    = \frac{q^2 \omega^2}{4 \pi^2}   | \vec{F}(\omega,\theta,\phi)|^2, 
    \label{FF}
    \end{equation} 
    with
    \begin{equation}
    \vec{F}(\omega,\theta,\phi) =  \hat{r} \times \int_{-\infty}^{\infty} d \lambda 
    \frac{d \vec{r}_q}{d \lambda} f(\lambda),
    \label{F}
    \end{equation}
    where
    $
    \vec{r}_q(\lambda) = 
    R\cos (\Omega \lambda) \hat{\rm i} + 
    R \sin(\Omega \lambda) \hat{\rm j} +
    a^{-1}\cosh (a \lambda) \hat{\rm k}
    $
    is the charge trajectory ($\{\hat{\rm i},\hat{\rm j}, \hat{\rm k}\} $ 
    being the usual Cartesian versors), 
    $\hat{r}$ gives the observation direction, 
    and
    $ f(\lambda) \equiv \exp 
    \left[ 
    -i \omega\left (\hat{r} \cdot \vec{r}_q (\lambda) - a^{-1} \sinh(a \lambda)\right) 
    \right] 
    $.
    The integrals in Eq.~(\ref{F}) can be solved by using 
    \begin{eqnarray}
    && \int_0^\infty d\zeta \, \zeta^{-im\Omega/a +\alpha- 1} 
    \exp{[({i \omega/(2a)}) \left(\Xi_- \zeta - \Xi_+ /\zeta  \right )]} 
    \nonumber \\
    && =2i^\alpha
    \exp\left[{\pi m\Omega}/{(2a)}\right]
    \left[\tan{(\theta}/{2})\right]^{-(im\Omega/a-\alpha)/2}
    \nonumber \\
    &&\times
     K_{im\Omega/a-\alpha} \left( \omega \sin \theta/a \right) 
    \end{eqnarray}
    for $\alpha= 0, \pm 1$ and $\Xi_\pm \equiv 1 \pm \cos \theta$~\cite{note4},
    combined with relations~\cite{note5}
    \begin{eqnarray}
    && \cos(\phi - \Omega \lambda) \exp{[-i k_\bot R \cos(\phi-\Omega \lambda)]} 
    \nonumber \\
    && = -\sum_m i^{m+1} J'_m(-k_\bot R) \exp{[i m (\phi-\Omega \lambda)]},
    \label{identidadeGabriel1}
    \end{eqnarray}
    \begin{eqnarray}
    &&R k_\bot \sin(\phi-\Omega \lambda) \exp{[-i k_\bot R \cos(\phi-\Omega \lambda)]} 
    \nonumber \\
    && =  \sum_m m i^m J_m(-k_\bot R) \exp{[i m (\phi-\Omega \lambda)]}.
    \label{identidadeGabriel2}
    \end{eqnarray} 
    Then, by using  Eq.~(\ref{FF}) into Eq.~(\ref{N}), we obtain 
    \begin{eqnarray}
     && \!\!\!\!\!\!\!\!        \frac{d^2 N_{k_\bot k_z}^{{\rm M} }}{dk_\bot d k_z } = \frac{2q^2 k_\bot }{\pi a^2 \hbar \omega } 
     \sum_{m=-\infty}^{\infty}  \exp ({{\pi m\Omega}/{a}})    ( |J_m(k_\bot R)|^2 
     \nonumber \\
     &&\!\!\!\!\!\!\!\! \times 
     |K'_{{i m \Omega}/{a}}\left({k_\bot}/{a} \right)|^2  \!\!+ \!
     (R \Omega)^2  |J'_m(k_\bot R)|^2   |K_{{i m \Omega}/{a}}\left({k_\bot}/{a} \right)|^2 ). \nonumber
     \end{eqnarray}

     Finally, by integrating it in $k_z$ and performing the redefinition  
     $k_z\to \kappa_z \equiv k_z/k_\bot$, we obtain
     \begin{equation}
     \frac{dN_{k_\bot}^{{\rm M} }}{dk_\bot} 
     =
     \left( \frac{4 \pi}{a} \int_{-\infty}^{\infty} \frac{d\kappa_z}{(1+\kappa_z^2)^{1/2}} \right)
     \left.\sum_{m=-\infty}^{\infty} \frac{d\Gamma_{k_\bot m}^{{\rm R}\, {\rm tot}}}{dk_\bot}\right|_{T=T_{\rm U}}.
     \label{final}
     \end{equation}
     This concludes our proof of Eq.~(\ref{Crucial}). The fact that the term between parentheses
     diverges is because the calculation above assumed a charge accelerating for infinite time, in 
     which case an infinite number of photons is emitted for fixed $k_\bot$ element. In real experiments 
     no divergence appears. We note that  Eq.~ (\ref{Crucial}) fits nicely real experiments with finite 
     windows provided $a_{\rm tot} \gg 1/L$~\cite{note_SM}. 

    {\bf Conclusions:}  We have proposed a simple experiment where the presence of the 
    Unruh thermal bath is codified in the Larmor radiation emitted from an accelerated 
    charge. Then, we carried out a straightforward classical-electrodynamics calculation 
    to confirm it by ourselves. Unless one challenges {\em classical} electrodynamics, 
    our results must be virtually considered as an observation of the Unruh effect. 

%%%%%%%%%%%%%%%%%%%%%%%%%%%%%%%%%%%%%%%%%%%%%%%%%%%%%%%%%%%%%%%%%%
    \acknowledgments
%%%%%%%%%%%%%%%%%%%%%%%%%%%%%%%%%%%%%%%%%%%%%%%%%%%%%%%%%%%%%%%%%%

    \textbf{Acknowledgments}: We are grateful to A.~J.~Roque da Silva and
    the Microtron group at the University of S\~ao Paulo for 
    explanations on electron accelerators.  G.~M.\ is indebted to A.\ Higuchi for
    various discussions. G.~C.\ and  A.~L., G.~M., D.~V.~were 
    fully and partially supported by S\~ao Paulo Research Foundation (FAPESP)
    under Grants 2016/08025-0 and 2014/26307-8, 2015/22482-2, 2013/12165-4,
    respectively. G.~M.\ was also partially
    supported by Conselho Nacional de Desenvolvimento
    Cient\'\i fico e Tecnol\'ogico (CNPq).

\section*{SUPPLEMENTAL MATERIAL: Alternative derivation of Eq.~(23) using standard quantum field theory}
Here we provide an alternative derivation of Eq.~(23) which reinforces that the infinite term between parenthesis appearing in this expression must be identified with the total Rindler proper time $\Delta \tau_{\rm R}$. In the inertial framework, the normalized physical modes of the  electromagnetic field in polar coordinates [solutions of Eq.~(6)] are
\begin{eqnarray}
    A_\alpha^{(I,m,k_\bot,k_z)} &=& k_\bot^{-1}
    \left( k_z g_{m k_\bot  k_z},-\omega g_{m k_\bot k_z} ,0,0\right), \nonumber
    \label{InertialObserversModes1} \\
    A_\alpha^{(II,m,k_\bot,k_z)} &=&  k_\bot^{-1}
    \left(0,0, -m g_{m k_\bot k_z} /r,  -i r \partial_r g_{m k_\bot k_z} \right), \nonumber
    \label{InertialObserversModes2}
\end{eqnarray}
where 
\begin{equation}
    g_{ m k_\bot k_z} \equiv 
    (8 \pi^2 \omega)^{-1/2} J_m \left( k_\bot r \right) e^{i m \phi} e^{i k_z z} e^{-i \omega t}, \nonumber
\end{equation}
$\epsilon = I, II$ labels the mode polarizations, $m \in \mathbb{Z}$, $k_\bot \in [0,+\infty)$, $k_z \in (-\infty,+\infty)$,  and we recall that $\omega = \left(k_\bot^2 +k_z^2 \right)^{1/2}$. We expand $\hat A_\alpha$ in terms of inertial normal modes as
\begin{equation}
    \hat A_\alpha = \sqrt{\hbar} \sum_{\epsilon,m} \int_0^{\infty} dk_\perp k_\perp \int_{-\infty}^{\infty} dk_z
    \left( \hat a_{(l)} A_\alpha^{(l)} + {\rm H.c.} \right)
    \label{expansion2} \nonumber
\end{equation}
with $(l)=(\epsilon, m, k_\bot, k_z)$. The $k_\bot$-distribution of Minkowski photons is given by
\begin{equation}
    k_\bot^{-1} \frac{dN_{k_\bot }^{{\rm M}\, {\rm }}}{dk_\bot}=
    \sum_{\epsilon,m}
    \int_{-\infty}^{\infty} d k_z 
    \ | {\cal A}^{{\rm M}\, {\rm em}}_{(l)}|^2,
    \label{emissionMrate}
\end{equation}
where
\begin{equation}
    |  {\cal A}^{{\rm M}\, {\rm em}}_{(l)} | = 
    \hbar^{-1} \left| 
    \int d^4 x \sqrt{-g}  j^\alpha\, 
    \langle l | \hat{A}_\alpha | 0 \rangle_M
    \right| \nonumber
\end{equation}
and we recall that $j^\alpha$ is given by Eq.~(4).
We note that Eq.~(\ref{emissionMrate}) [in contrast to Eqs. (10) and (11)] does not carry any thermal factor because the Minkowski vacuum, $|0 \rangle_M$, is a no-particle state according to inertial observers. The photon emission amplitudes for both polarizations can be written as
\begin{eqnarray}
    {\cal A}^{{\rm M}\, {\rm em}}_{(I,m,k_\bot,k_z)}  &=& \hbar^{-1}\langle  I , m,   k_\bot,  k_z | i S_I | 0 \rangle_{M} \nonumber \\
    &=& \frac{i q J_m \left(k_\bot R \right)}{2 \pi  k_\bot \sqrt{2 \hbar \omega}}   \int_{-\infty}^{\infty} d\lambda \  h(\lambda)  \nonumber \\
    &\times& \left[ k_z \cosh{(a \lambda)}  - \omega \sinh{(a \lambda)}\right], 
    \label{Ampl1} \\  
    \nonumber \\     
    {\cal A}^{{\rm M}\, {\rm em}}_{(II,m,k_\bot,k_z)} &=& \hbar^{-1} \langle II, m, k_\bot, k_z  | i S_I | 0 \rangle_{M} \nonumber \\
    &=&\frac{ q R \Omega J'_m(k_\bot R) }{2 \pi   \sqrt{2 \hbar \omega}}
    \int_{-\infty}^{\infty} d\lambda \  h(\lambda), 
    \label{Ampl2}
\end{eqnarray}  
where
\begin{equation}
    h(\lambda) = \exp{\left(-i m \Omega \lambda -\frac{i}{a} \left[ k_z \cosh{(a \lambda)}-\omega \sinh{(a \lambda)} \right] \right)} \nonumber
\end{equation}
and $\lambda$ is related to the inertial time by $t = a^{-1} \sinh{(a\lambda)}$, i.e., $\lambda$ is the proper time of the Rindler observer at $\xi=0$. This variable change is a necessary maneuver to allow us to express the emitted photon number in terms of $\Delta \tau_{\rm R}$.   

In order to obtain the total emitted photon number per fixed $k_\bot$, we must square the absolute values of the amplitudes~(\ref{Ampl1}) and~(\ref{Ampl2}) and insert them in Eq.~(\ref{emissionMrate}). From this procedure, we end up with an integral 
$\int_{-\infty}^{\infty} d\lambda \int_{-\infty}^{\infty} d\lambda' \ (\cdots)$ which can be expressed 
as $\int_{-\infty}^{\infty} d\tau_{\rm R} \int_{-\infty}^{\infty} d\sigma \ (\cdots)$ once we write $\tau_{\rm R} = (\lambda+\lambda')/2$ and $\sigma = \lambda-\lambda'$. As a result, we obtain for the emitted photon number
\begin{eqnarray*}
    &&k_\bot^{-1} \frac{dN_{k_\bot }^{{\rm M}\, {\rm}}}{dk_\bot} =  
    \Delta \tau_{\rm R} \sum_m  \frac{q^2}{8 \pi^2 \hbar}
    \int_{-\infty}^{\infty} d\sigma \int_{-\infty}^{\infty} d k_z  \\
    &&  \times  \left[ \left( {\cal N}_I + {\cal N}_{II} \right)  \exp [-i m \Omega  \sigma - 2i a^{-1} \omega \sinh{\left(a \sigma/2 \right)}] \right],
\end{eqnarray*}
with
\begin{eqnarray*}
    \Delta \tau_{\rm R} &=& \int_{-\infty}^{\infty} d\tau_{\rm R}, \\
    {\cal N}_I &=&  \frac{|J_m(k_\bot R)|^2}{\omega} \left[\left(\frac{\omega}{k_\bot} \right)^2-\cosh^2 \left({\frac{a \sigma}{2}} \right) \right], \\
    {\cal N}_{II} &=& \frac{|J'_m (k_\bot R)|^2}{\omega} (R \Omega)^2 ,
\end{eqnarray*}
where we have made the transformation $(k_z,\omega) \to (\tilde{k}_z,\tilde{\omega})$ with
\begin{eqnarray*}
    \tilde{k}_z & \equiv& k_z \cosh(a \tau_{\rm R}) - \omega \sinh(a \tau_{\rm R}), \\
    \tilde{\omega} & \equiv&  \omega \cosh(a \tau_{\rm R}) - k_z \sinh(a \tau_{\rm R}),
\end{eqnarray*}
and dropped the tildes, eventually. After solving the remaining integrals in $\sigma$ and $k_z$, we obtain
\begin{eqnarray}
    && \frac{dN_{k_\bot }^{{\rm M}}}{dk_\bot} 
    = \Delta \tau_{\rm R} \frac {q^2 k_\bot}{ \pi^2 \hbar a}\sum_m \Theta(m)  \left[ |K'_{{im\Omega}/{a}} ({k_\bot}/{a})|^2 \right.
    \nonumber \\
    && \left. \times   | J_m (k_\bot R ) |^2 + (R \Omega)^2 | K_{{i m \Omega}/{a}} ({k_\bot}/{a} ) |^2 | J'_m (k_\bot R) |^2  \right]
    \nonumber \\
    && \times \cosh \left[ \frac{\pi m \Omega}{a}\right],
\end{eqnarray} 
which coincides with Eq.~(23) provided we make the identification
\begin{equation*}
    \Delta \tau_{\rm R} \leftrightarrow  \left( \frac{4 \pi}{a} \int_{-\infty}^{\infty} \frac{d \kappa_z}{\left( 1+\kappa_z^2\right)^{1/2}} \right).
\end{equation*}


\begin{thebibliography}{}
\bibitem{Note_velocities}
    Setting $Z\equiv z-a^{-1} = 0$ at the point where the charge makes the ``U turn'' in the longitudinal
     direction ($z$) and adopting orientation such that $E^z<0$, the injection
    velocity components are given by
    $v_{\bot} = \Omega R /(1+a Z_i)$ and $v_\parallel = -\sqrt{a^2 Z_i^2+ 2 a Z_i}/(1+a Z_i)$,
    where $Z_i>0$ is the position where the injection occurs.
\bibitem{note6}
    For  wavelengths $\lambda$ larger than the bunch size, interference 
    plays an important role and the whole set of charges acts collectively 
    as a single one with magnitude $q$ equal to the sum of all charges. This magnifies 
    the emitted power which scales as $q^2$.
\bibitem{note7}
    Using Larmor formula, we see that for these parameters, the charged bunch will 
    emit $\sim 10 \times L/(1~{\rm m})~{\rm GeV}$, which is small
    even when compared to the fraction of the total kinetic energy carried in the
    form of rotational energy of the bunch, $\sim 10^4~{\rm GeV} $.
\bibitem{note3}
    We note, in addition, that due to angular momentum conservation, the absorption of a  photon with $m>0$ 
   by 
    the charge as seen by accelerated observers corresponds to the emission of 
    a photon with $m<0$ according to inertial ones.
\bibitem{note4}
    This expression can be obtained from Eq.~8.432.7 of 
    Ref.~\cite{GR} after the substitution $e^{a\lambda} \to \zeta$ 
    and ``rotation'' to the positive imaginary axis.
\bibitem{note5}
    These identities are obtained by properly differentiating Eq.~8.511.4 
    of Ref.~\cite{GR} (after the substitution $z\to -\omega R \sin \theta$ 
    and $\varphi \to \phi-\Omega \lambda$).
\end{thebibliography}
    \end{document}